\documentclass[twocolumn]{ws-jai}
\usepackage{graphicx,amssymb,hyperref,nicefrac,xcolor}

\def\simgt{\lower.5ex\hbox{$\; \buildrel > \over \sim \;$}}
\def\etal{{\it et al.}}

\def\rmd{\mathrm{d}}

\def\lconfocal{\ell_\mathrm{confocal}}
\def\lspline{\ell_\mathrm{spline}}
\def\bx{\mathbf{x}}

\newcommand{\be}{\begin{equation}}
\newcommand{\ee}{\end{equation}}
\newcommand{\ba}{\begin{eqnarray}}
\newcommand{\ea}{\end{eqnarray}}

\newcommand{\cfai}{1}
\newcommand{\icc}{2}
\newcommand{\jpl}{3}
\newcommand{\toronto}{4}
\newcommand{\edi}{5}

\begin{document}

\catchline{}{}{}{}{} % Publisher's Area please ignore

\markboth{Clark et al.}{Characterization of an interline transfer CCD camera}

\title{Characterization of a commercial, front-illuminated interline transfer\\CCD camera for use as a guide camera on a balloon-borne telescope}

\author{
Paul Clark$^{\cfai}$,
Richard Massey$^{\cfai,\icc}$,
Herrick L.\ Chang$^{\jpl}$,
Mathew Galloway$^{\toronto}$,
Holger Israel$^{\icc}$,
Laura L.\ Jones$^{\jpl}$,
Lun Li$^{\toronto}$, \\
Milan Mandic$^{\jpl}$,
Tim Morris$^{\cfai}$,
Barth Netterfield$^{\toronto}$,
John Peacock$^{\edi}$,
Ray Sharples$^{\cfai}$ and
Sara Susca$^{\jpl}$}

\address{
$^{\cfai}${Centre for Advanced Instrumentation, University of Durham, Joseph Swan Road, Netpark, Sedgefield TS21 3FB, United Kingdom}\\
$^{\icc}${Institute for Computational Cosmology, University of Durham, South Road, Durham DH1 3LE, United Kingdom}\\
$^{\jpl}${Jet Propulsion Laboratory, California Institute of Technology, 4800 Oak Grove Drive, Pasadena, CA 91109, United States}\\
$^{\toronto}${University of Toronto, Department of Physics, 60 St.\ George St., Toronto, Ontario M5S 1A7, Canada}\\
$^{\edi}${Institute for Astronomy, University of Edinburgh, Royal Observatory  Blackford Hill, Edinburgh, EH9 3HJ, United Kingdom}\\}

\twocolumn[{

\maketitle

\begin{history}
\received{June 27, 2014};
\revised{August 26, 2014};
\accepted{August 27, 2014}.
\end{history}

\begin{abstract}
We report results obtained during the characterisation of a commercial front-illuminated progressive scan interline transfer CCD camera.
We demonstrate that the unmodified camera operates successfully in temperature and pressure conditions ($-40^\circ$C, $4$\,mBar) representative of a high altitude balloon mission.
We further demonstrate that the centroid of a well-sampled star can be determined to better than 2\% of a pixel, even though the CCD is equipped with a microlens array.
This device has been selected for use in a closed-loop star-guiding and tip-tilt correction system in the BIT-STABLE balloon mission.
\end{abstract}


}]

\section{Introduction}
\label{sect:intro}

High altitude balloons offer a revolutionary environment to conduct optical imaging and spectroscopy for astronomy and Earth observation. 
From an altitude of 35\,km (above 99\% of the Earth's atmosphere), image quality potentially equals that of satellites in near-Earth orbit, but at a fraction ($\sim$1\%) of the cost and with the opportunity to repair or upgrade hardware between flights (Rhodes et al.\ 2012). 
Furthermore, the recent development of Ultra Long Duration, super pressure balloons (ULDBs) allows large surveys during flights that last several months.

Balloon missions face limited budgets on cost and complexity (and size and mass).
Commercial, off-the-shelf hardware can be adopted to stay within these budgets, but only if its performance meets %scientific 
requirements.
In this paper, we describe tests of an Imperx Bobcat IGV-B2320-M % camera, validating its ability 
to (i) operate in a flight-like temperature and pressure environment (-40$^\circ$C, 4\,mBar) and (ii) perform science-quality imaging.
This camera was one of several candidates considered for the star-tracking guide camera of the BIT-STABLE project, a collaboration between NASA/JPL, the University of Toronto, and a UK consortium (Durham University and the University of Edinburgh) to fly an optical telescope on a high altitude balloon.

The Imperx camera is available with an ethernet interface that simplifies interfacing to the flight computer and makes a flight-compatible frame-grabber unnecessary.
Its internal electronics have already been ruggedized for use in industrial settings, for example with a thermally-optimised (fan-less) design and internal conformal coating.
We use an environmental chamber to recreate the temperature and pressure cycle of a typical balloon mission, and explicitly check that the camera will both survive and continue to operate.

The imaging detector is a Truesense KAI-04050 front-illuminated interline transfer CCD, with $2352\!\times\!1768$ 5.5\,$\mu$m (square) pixels and 12-bit readout.
Interline transfer CCDs have a nonresponsive area in each pixel, which is used for fast readout.
This can usefully mimic an electronic shutter, reducing complexity.
However, a lenslet array is then needed to focus incident light onto the responsive area in each pixel to restore QE, {and distortion through these optical elements may spuriously move the apparent position of a guide star}.
Using a precision translation stage, we move a spot, representative of the BIT-STABLE PSF, within several of the device's pixels.
We then explicitly compare the known position of the spot to the position of the pixel in which light is actually registered.
During testing, we also noticed anomalies in images obtained in Windowed Readout Mode that could also affect position measurements.

This paper is organized as follows.
In section~\ref{sec:cryo}, we describe camera operational tests in an environmental chamber that replicates flight conditions.
In section~\ref{sec:cent}, we describe measurements of sub-pixel variations in detector response. 
In section~\ref{sec:window}, we describe measurements of anomalies in windowed captures, investigate their possible origin, and develop a mitigation strategy.
We conclude in section~\ref{sec:conc}.

\section{Operation under flight conditions}
\label{sec:cryo}

\begin{figure}[t]
\centering
\includegraphics[width=85mm]{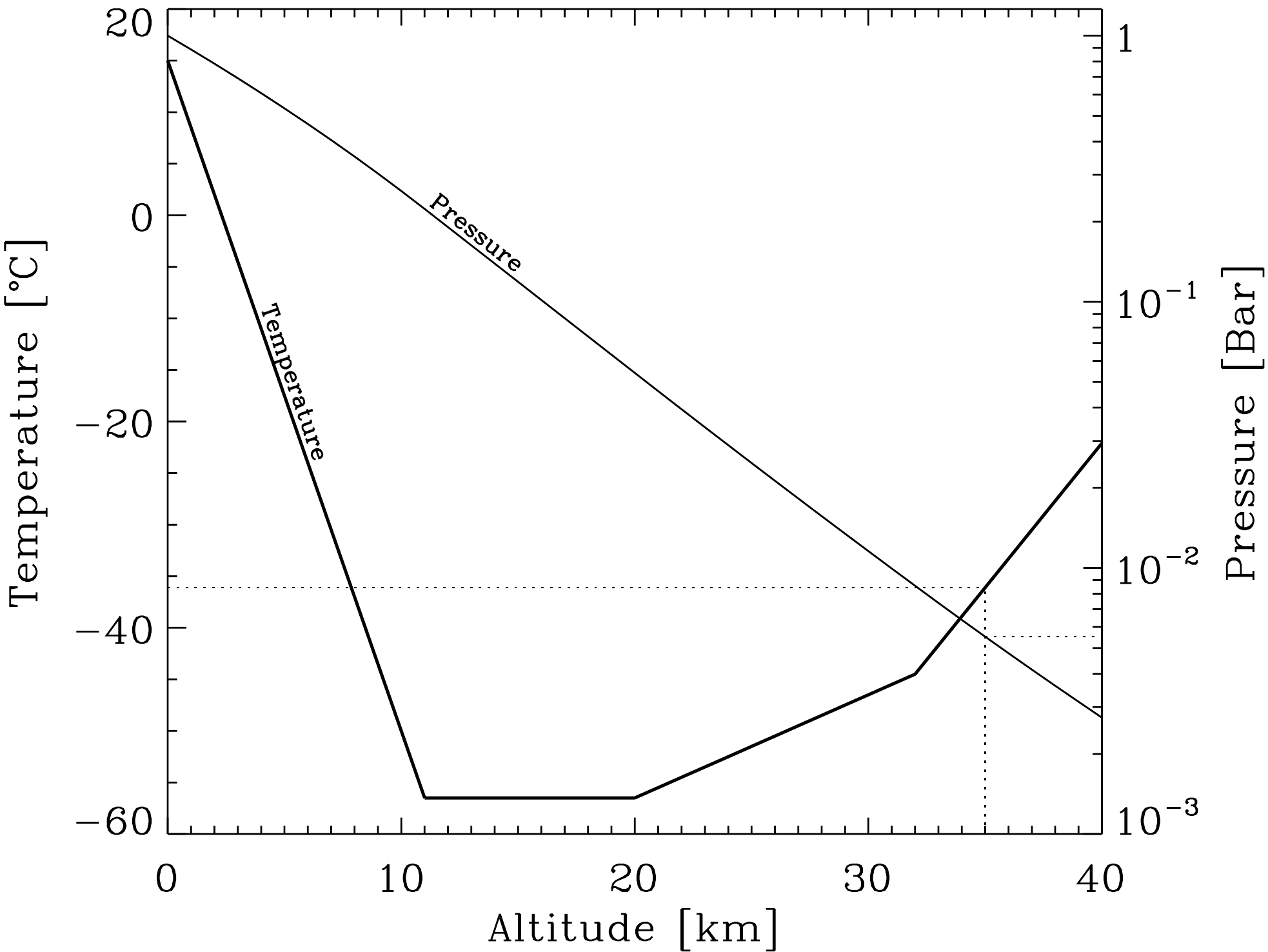}
\caption{Environmental conditions as a function of altitude, in the 1976 U.S.\ Standard Atmosphere model.}
\label{fig:environment}
\end{figure}

\subsection{The problem}
\label{sect:cond-problem}

The BIT-STABLE balloon mission will operate at an altitude of 35\,km.
As seen from the U.S.\ Standard Atmosphere \citep[see figure~\ref{fig:environment}; ][]{us_standard_atmos}, the ambient pressure and temperature at that altitude are approximately $-40^\circ$C and 4\,mBar (0.004\,atm). 
The pressure falls exponentially during ascent, while the temperature profile passes through a minimum of $-56^\circ$C at an altitude of 11--20\,km before rising again towards float altitude.
The ascent to 35\,km is expected to take approximately 100\,minutes (based on balloon flight data from BLAST; \citealt{blast06}), so the balloon hardware will only encounter ambient temperatures as low as $-56^\circ$C for a relatively short time. 
Since the telescope structure and gondola provide significant thermal mass and some heating, the BIT-STABLE mission has defined requirements of $-40^\circ$C survivability temperature and $-30^\circ$C minimum operating temperature for all flight hardware. 

As the air pressure decreases with altitude, convective cooling also decreases, perversely resulting in the flight hardware being able to overheat rather than becoming too cold. 
BLAST used air-filled pressure vessels to maintain convective cooling of flight electronics \citep{blast};
BIT-STABLE's financial and weight budgets (the camera must be moved to track the sky) make pressure vessels undesirable. 
It is therefore essential to test candidate electronics, including the guide camera, under representative flight conditions to ensure they can operate reliably in the low temperature, low pressure, low convective cooling environment.

The Imperx Bobcat range of cameras are available with an internal conformal coating and have a thermally optimised, fan-less design allowing them to operate over an extended temperature range of $-40^\circ$C to $+85^\circ$C (in air). 
Operation at high altitude should be possible so long as the $\sim$5\,W internal power dissipation doesn't raise the internal temperature to unsafe levels, or the low pressure cause failure of (e.g.) electrolytic can capacitors. The project team at Durham University therefore subjected an IGV-B2320 to a series of environmental tests prior to shipping to JPL. Initial tests concentrated on establishing the level of self-heating in the camera at reduced pressure. Later tests established the camera's ability to survive and operate in the low pressure and temperature environment.

\subsection{Experimental setup}
\label{sect:cond-exp}

\subsubsection{Experimental configuration}

\begin{figure}[t]
\centering
\includegraphics[width=85mm]{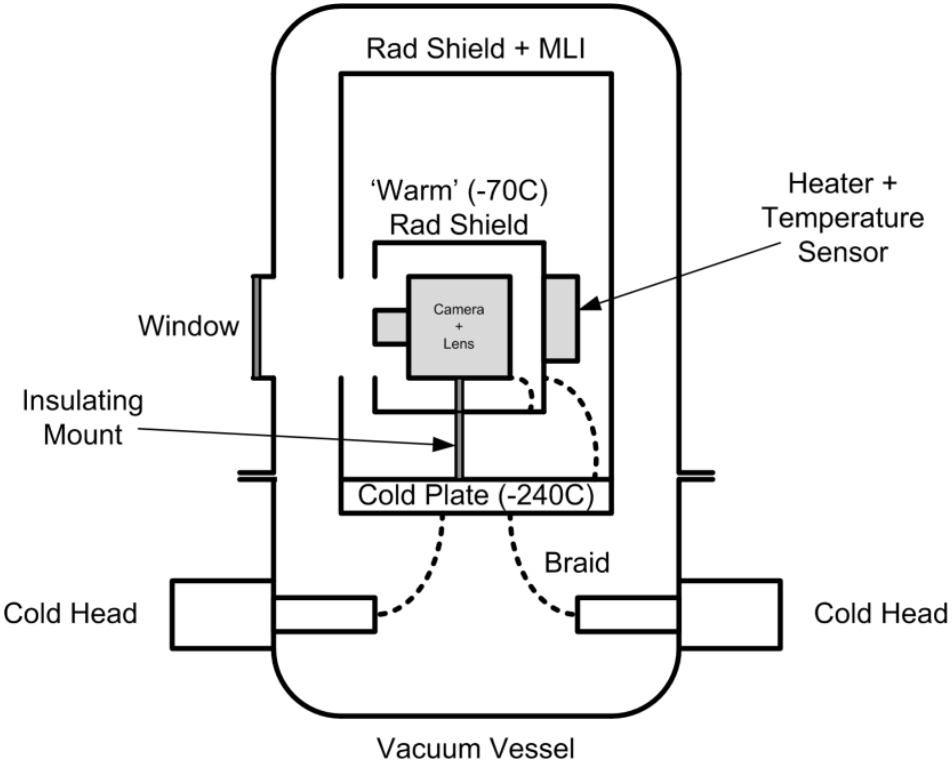}
\caption{
Diagram of components inside the cryogenic chamber used to test camera survivability and operability at flight temperature and pressure.
}
\label{fig:cryochamber}
\end{figure}

\begin{figure}
\centering
\includegraphics[width=84mm]{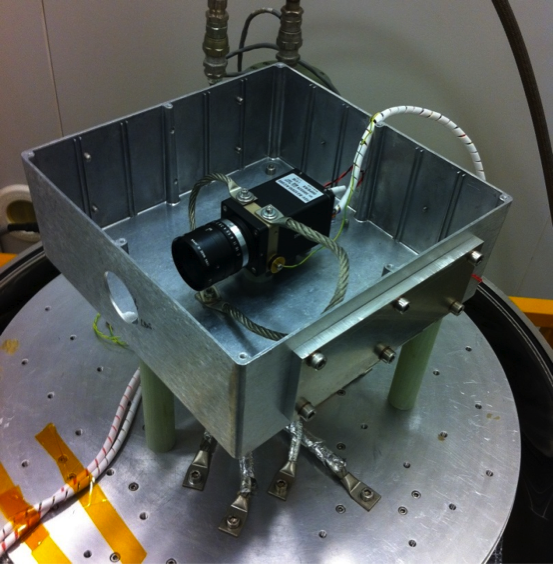}
\caption{
Photograph of the camera inside the secondary radiation shield (see figure~\ref{fig:cryochamber}), with the lid removed.
A pair of internal thermal braids connects the camera to the radiation shield; these were absent during initial runs to measure self-heating.}
\label{fig:cryocamera}
\end{figure}

\begin{figure}
\centering
\includegraphics[width=84mm]{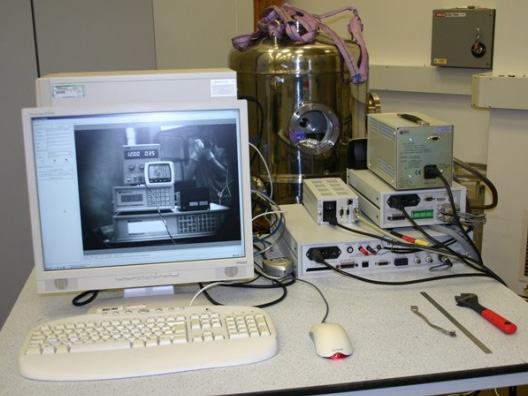}
\caption{Photograph of the closed cryogenic chamber (see figure~\ref{fig:cryochamber}) and support equipment. 
The monitor shows an image taken by the camera looking out from inside the chamber, showing the power supply and Lakeshore controller etc.}
\label{fig:cryocontroller}
\end{figure}

To replicate flight conditions in the laboratory, a bespoke cryogenic test chamber was used, as illustrated in figures~\ref{fig:cryochamber}--\ref{fig:cryocontroller}.
The test chamber contains a 50\,cm diameter cold bench cooled by two CTI Cryo-Torr cold heads.
A cylindrical radiation shield and multi-layer insulation (MLI) prevent the cold mass from being warmed by radiation from the room temperature vessel. 
During operation the cold bench stabilises at a temperature of approximately $-240^\circ$C, much too cold for the project's needs. 
The Imperx camera was therefore mounted inside a secondary radiation shield, an aluminium alloy box, connected to the cold bench by thermal braids.
The cross-sectional area of the cold braids between the box and the cold bench were tuned to ensure the box would be cooled to approximately the desired temperature, then more finely regulated by two polyimide film heaters mounted on the box and controlled by an external Lakeshore temperature controller.

Initial tests were conducted with the camera isolated from the box, to determine the level of self-heating when operating in near-vacuum. 
For subsequent tests, small diameter cold braids (visible in figure~\ref{fig:cryocamera}) were installed between the camera and the box. 
By regulating the temperature of the box to approximately $-70^\circ$C, the camera temperature could be lowered to the required $-30^\circ$C (operating) or $-40^\circ$C (non-operating, survival). 
Vacuum feed-through connectors allowed the camera power and ethernet connections to be brought to the outside. A hole was cut in the radiation shield to allow the camera to take images through the chamber window, of the Lakeshore controller, Lakeshore temperature monitor and Leybold vacuum gauge located adjacent to the chamber (figure~\ref{fig:cryocontroller}). Temperatures were recorded via Lakeshore diode sensors. An analogue wallclock provided a useful time reference when the images taken periodically by the camera were turned into a time lapse movie sequence.

\subsubsection{Camera GigE interface}\label{sec:gige}

On three separate computers (2 Windows, 1 Linux), the camera's Gigabit ethernet connection would occasionally produce dropped frames or partial images. 
We postulate that there is a certain dependence either on the ethernet drivers or ethernet hardware that is not fully compatible with the GigE Vision standard.
By forcing the computers' connection back to 100~Mbps Full Duplex mode, this anomaly disappeared. 
This workaround will bottleneck the maximum frame rate for large windows or full frame captures.
However, occasional full frame readout, or $50$\,Hz readout of windows smaller than $\sim\!256\!\times\!256$\,pixels, remains possible within the 100~Mbps ethernet limit.

\subsection{Results}
\label{sect:cent-results}

\subsubsection{Self-heating}

\begin{figure}[t]
\centering
\includegraphics[width=85mm]{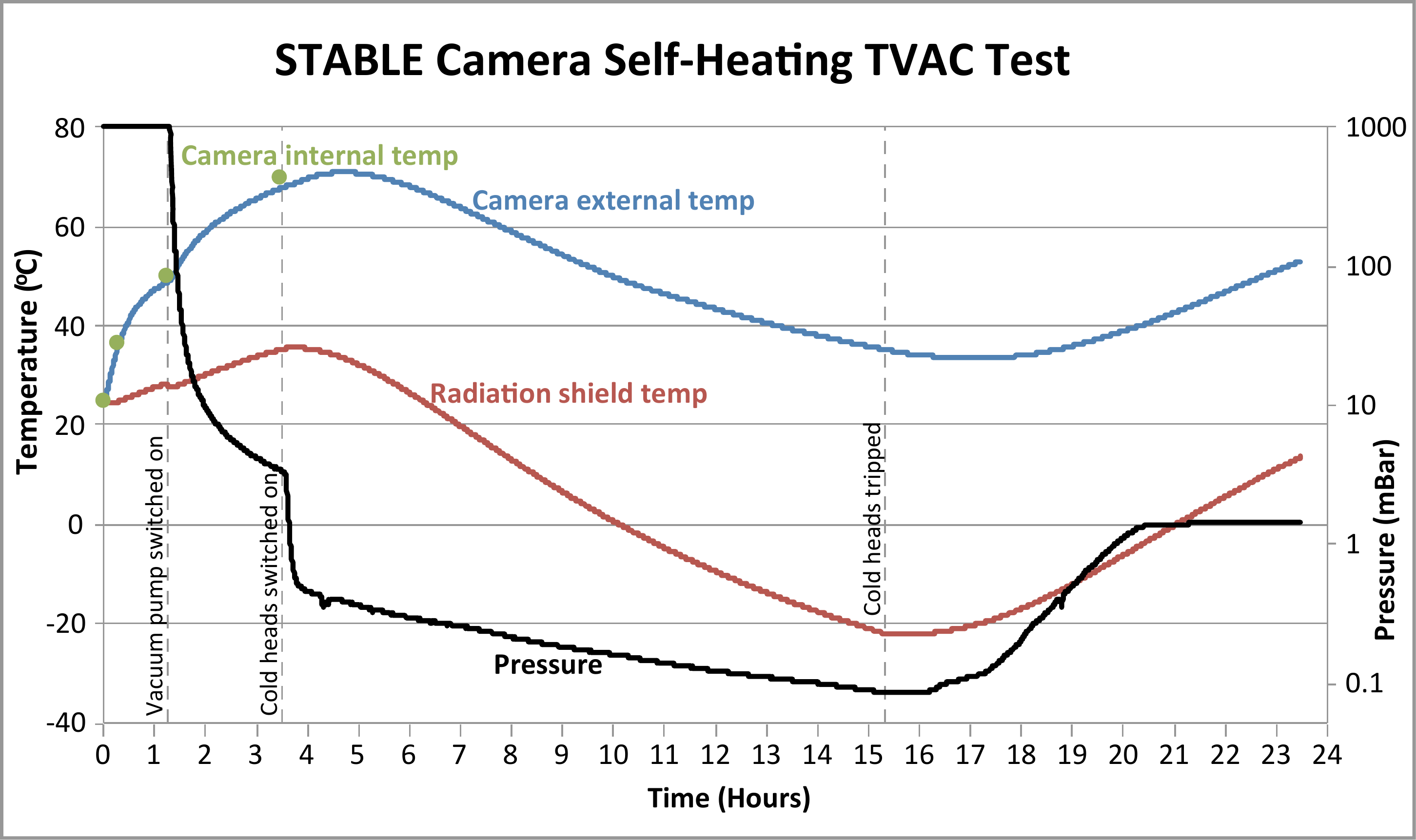}
\caption{Temperature and pressure profile during initial thermal cycle to measure self-heating (without cold braids connected to the camera).}
\label{fig:cryo_self}
\end{figure}

To measure the level of self-heating in the camera, cold braids between the camera and the secondary radiation shield were omitted for an initial thermal cycle.
The camera was operated continuously throughout the test, with full frame images recorded every minute (later to be turned into a time-lapse sequence), although power consumption of the camera did not vary measurably if images were taken at a higher frame rate.
The recorded temperature and pressure profiles are shown in figure~\ref{fig:cryo_self}.
\begin{itemize}
\item
T=0: the camera was powered up, in atmosphere. The temperature of the camera body, recorded via a Lakeshore temperature monitor, rose above the $22^\circ$C ambient to $50^\circ$C. The camera internal temperature, reported by the GEV software, was logged occasionally and observed to stabilise at approximately $2^\circ$C above the temperature of the camera body --- indicating good thermal contact between the two.
\item
T=1.25\,hours: the vacuum pump was switched on and the pressure inside the vacuum chamber was seen to fall. The vacuum pump, a large dry scroll pump, was connected to the chamber via 6\,mm plastic hose, rather than conventional large bore vacuum hose, to deliberately reduce the decrease in pressure to a rate similar to what the camera will encounter during flight. The camera exhibits a corresponding increase in temperature as the convective cooling is reduced.
\item
T=3.4\,hours: the cold heads were switched on when the camera reached a maximum temperature of $70^\circ$C. With the cold heads running, the pressure is seen to drop more rapidly as cryo pumping takes place. This is unavoidable and unfortunately takes the camera below the operational pressure of 4\,mBar. The temperature of the camera was observed to fall, matching the fall in the temperature of the secondary radiation shield.
\item
T=15.3\,hours: the cold heads were switched off causing the temperatures and pressure to rise accordingly.
\end{itemize}

The camera stabilises at $\sim\!45^\circ$C above ambient in vacuum without cold braids, due to $\sim$5\,W internal power dissipation. If the camera were mounted on an insulating mount on the BIT-STABLE telescope, we could therefore expect the camera temperature to remain above $0^\circ$C throughout the flight: reducing the risk of frosting during ascent or decent. However, a desire to minimise dark current in the CCD may require the camera to be thermally strapped to the cold mass of the telescope.

\subsubsection{Survivability and operational tests}

\begin{figure}[t]
\centering
\includegraphics[width=85mm]{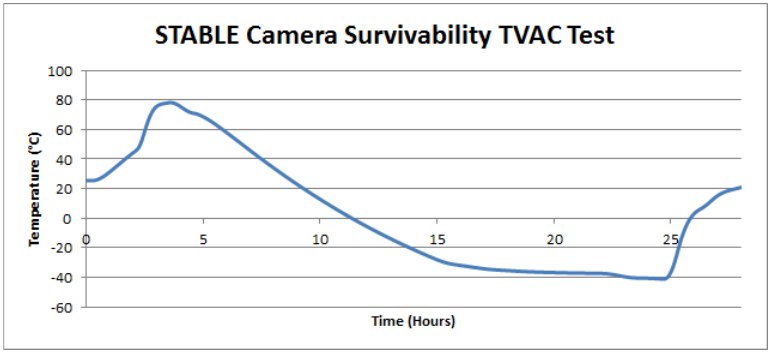}
\includegraphics[width=85mm]{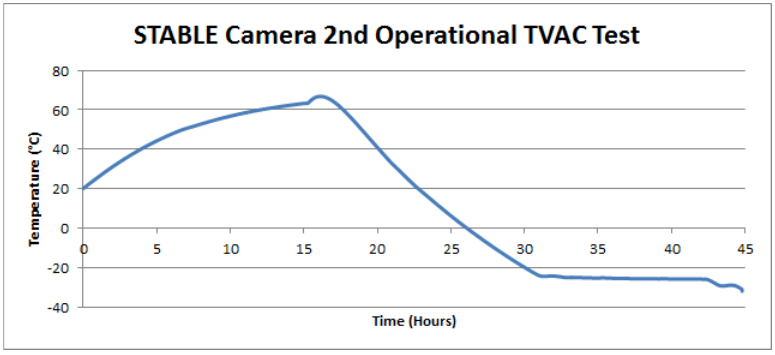}
\caption{Temperature profile during survivability and operational tests of the camera (with cold braids connected to the camera, enabling it to be sufficiently cooled).}
\label{fig:survivability}
\end{figure}

For the subsequent survivability and operational tests, cold braids were installed between the camera body and the inner radiation shield, as shown in figure~\ref{fig:cryocamera}.
These allowed the camera to be cooled to the required $-40^\circ$C survivability and $-30^\circ$C operating temperatures.
The survivability test required the camera to be warmed to $+70^\circ$C, dwell for one hour, then cooled to $-40^\circ$C, dwell for one hour, and then return to ambient. 
The camera was not powered during this test. 
The operational test required the camera to be warmed to $+60^\circ$C, dwell for one hour, then cooled to $-30^\circ$C, dwell for one hour, and then returned to ambient. 
The camera was continuously operational throughout this test. The operational test was repeated a second time. 
Figure~\ref{fig:survivability} shows the camera body temperature profiles during the survivability and second operational tests.

As the camera under test would be used for flight, it was necessary to avoid subjecting it to any unnecessary temperature cycles. 
This meant that setting the temperature of the secondary radiation shield to achieve the desired camera temperature involved a little educated guesswork. 
With repeated testing the target temperatures could be determined more accurately, but for the first time tests shown in figure~\ref{fig:survivability}, the temperature profiles are certainly non-optimal. 
During the survivability test for example, the secondary radiation shield temperature was initially set a little above $+70^\circ$C. 
After 2.5 hours it became clear that the (non-operating) camera was going to take a long time to reach the required $+70^\circ$C, so the radiation shield temperature was raised. 
This then resulted in the camera temperature overshooting and reaching close to $+80^\circ$C. 
Having been above $+70^\circ$C for the required one hour, cooling was started and the secondary radiation shield temperature was set to a little below $-40^\circ$C. 
At T=22 hours, the following morning, the camera had not quite reached $-40^\circ$C, so the radiation shield temperature was lowered. 
The camera then spent the required one hour at $-40^\circ$C before the cooling was switched off and the camera warmed back to ambient. 
The profile for the operational test is similar with the radiation shield temperatures having to be nudged to achieve the target $+60^\circ$C / $-30^\circ$C (operating) camera temperature.

The camera passed the survivability and both operational tests, and continues to function well during integration into the telescope system at JPL.

\section{Sub-pixel Photo Response Non-Uniformity}
\label{sec:cent}

\subsection{The problem}
\label{sect:cent-prob}

The architecture of an interline CCD is ideal for a balloon telescope's guide camera in several ways.
At the end of an exposure, charge is stepped sideways by one pixel, into a column that is non-responsive to light but dedicated to charge transfer and readout.
This architecture avoids the need for a mechanical shutter, which would otherwise be both a single point of failure and a source of vibration.
It allows the next exposure to begin almost immediately, optimizing frame rate and exposure time.
It also avoids image smearing in the readout direction that would be present in even a frame-transfer CCD when the frame rate is high, and the exposure time approaches the transfer time.

However, since part of the surface of an interline CCD surface is non-responsive, and this device is front-illuminated, its Quantum Efficiency (QE) is low.
To compensate, a grid of microlenses is incorporated into the CCD surface, which focuses incident light into the responsive regions.
This raises the QE to 47\% at 510\,nm \citep{truesense}.
Unfortunately, this extra optical surface (and the electrode structure in a front-illuminated CCD) raises the potential for additional sources of detector Photo Response Non-Uniformity (PRNU) on sub-pixel scales: {in some locations, light may be deflected toward the wrong pixel}, or scattered and lost.
This is of particular concern because the 
{particular flowdown and budget allocation of science requirements in the BIT-STABLE mission means that the guide camera must enable position measurements with an} accuracy better than 250\,nm=0.045\,pixels.
To assess this camera's performance, we scan a small spot across the detector, recording its true position and comparing this to its measured position (and flux).

\subsection{Experimental setup}
\label{sect:cent-exp}

\begin{figure}[t]
\centering
\includegraphics[width=85mm]{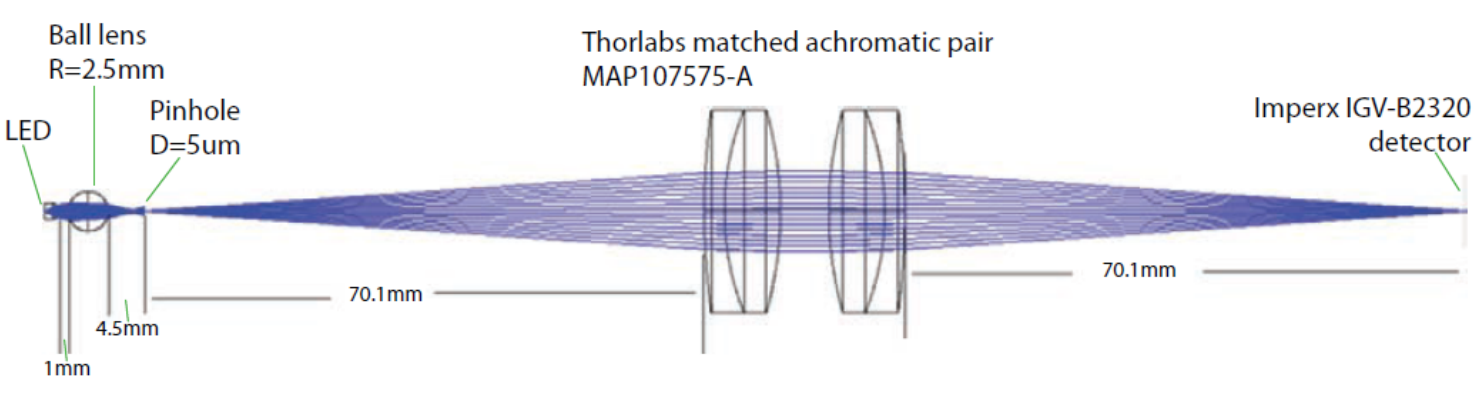}
\includegraphics[width=85mm]{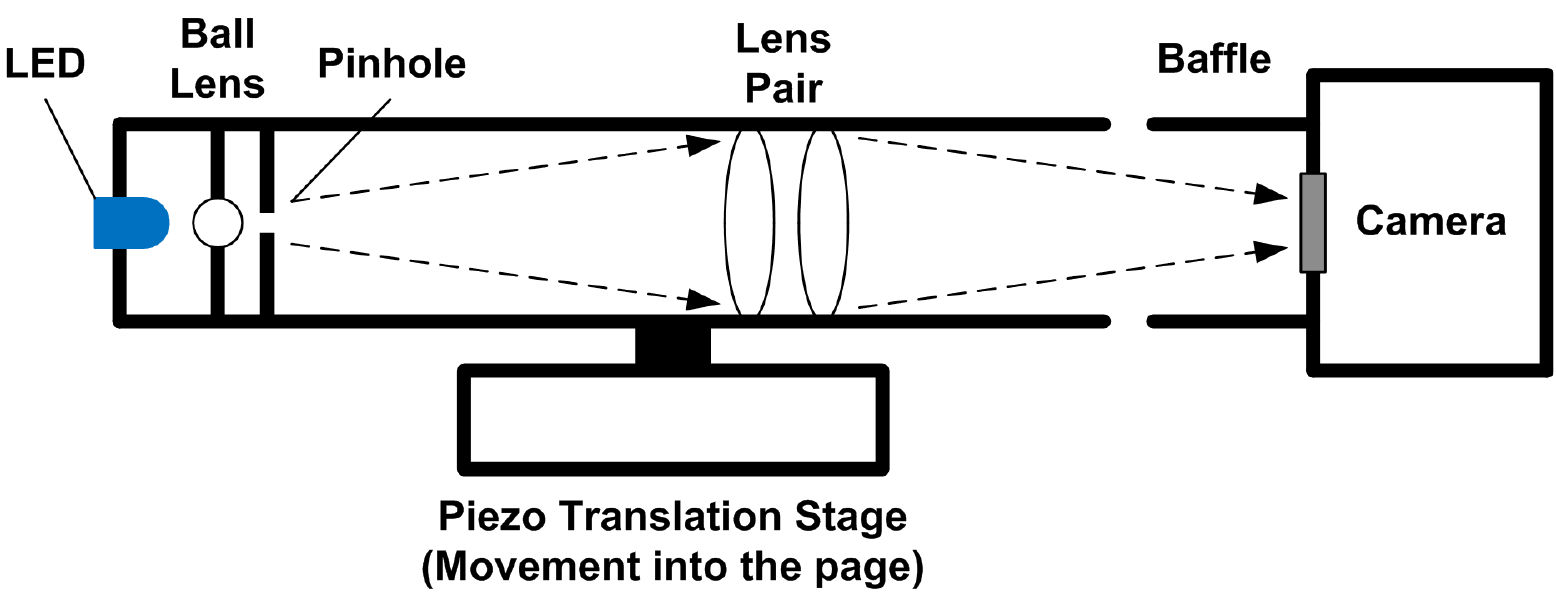}
\caption{Zemax model and schematic diagram of components used to shine a spot towards the camera, and move it around the detector at high precision.}
\label{fig:lab_setup_diagram}
\end{figure}

To test the centroid determination accuracy, an illuminated pinhole was re-imaged onto the 
CCD surface. A piezo translation stage moved the pinhole source over the CCD surface with 
high positional accuracy. A separate confocal probe was used as a truth sensor, to independently monitor the stage position.
Figures~\ref{fig:lab_setup_diagram}--\ref{fig:lab_setup2} show the experimental set-up.

The pinhole source was constructed using Thorlabs SM1 components. 
The interchangeable 5mm LED was held in a S1LEDM mount. 
A $5\,$mm BK7 ball lens, held in a customised SM1PL plug, focussed the light from the LED onto the pinhole (e.g.\ P5S). 
An SM1V10 focussing tube plus SM1L05 and SM1L10 lens tubes set the separation between the pinhole and the lens. 
A MAP107575-A matched achromat lens pair ($75\,$mm focal length) with AR coating re-imaged the pinhole $1\!:\!1$ onto the CCD surface. 
The lens pair has a quoted RMS spot size of $7.7\,\mu$m. 
The default setup used a $5\,\mu$m diameter pinhole and an LED with a central wavelength of $472\,$nm.

An LM371L $5\,$mA constant current source powered the LED. 
Thin enamelled copper wires were used to make the final connections to the LED to avoid restricting the piezo stage movement. 
A PICAXE 20X2 micro-controller and ULN2803A driver allowed the LED to be turned on and off via USB under the control of Python software running on a Windows PC. 
A Physik Instrumente (PI) P-625.1CD piezo translation stage ($500\,\mu$m) was used to move the pinhole source over the CCD surface. 
The stage was operated in closed-loop with capacitive feedback. 
The $170\,$g mass of the pinhole source required the $P$, $I$, slew rate and notch filter parameters to be configured carefully to avoid causing oscillations in the stage position, and the optics were cushioned to damp any oscillations that did arise. 
The stage used for this experiment did not have digital dynamic linearization and so it was necessary to monitor the stage position with a separate Micro Epsilon confocal probe ($300\,\mu$m) to allow the residual micron-level ripple in the stage linearity to be removed during the centroid calculation. 

\begin{figure}
\centering
\includegraphics[width=85mm]{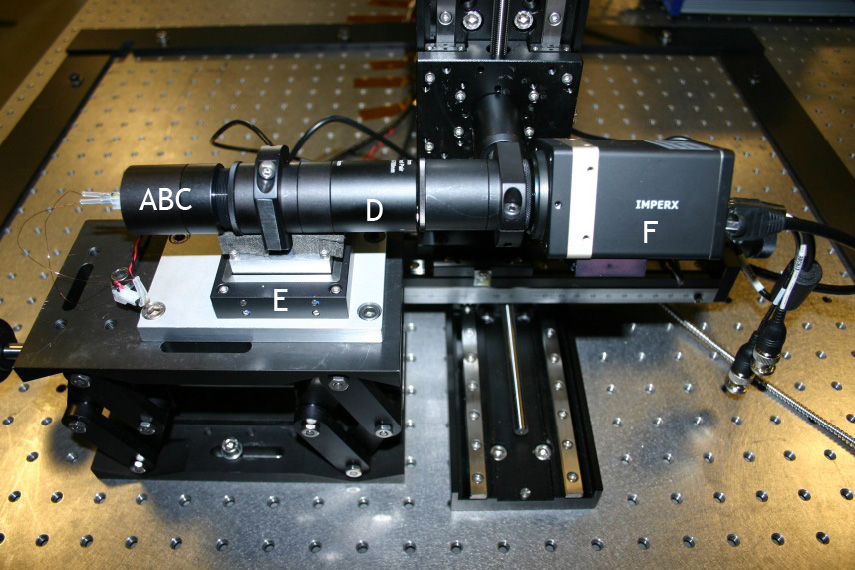}
\caption{Experimental set-up to scan a spot across the detector.
Side view (as in figure~\ref{fig:lab_setup_diagram}). Key: A: 5mm LED; B: 5mm BK7 ball lens (not shown); C: mounted pinhole (not shown); D: 1:1 matched achromatic doublet pair; E: piezo translation stage; F: CCD camera with baffle tube. Confocal probe removed.}
\label{fig:lab_setup2}
\end{figure}

\begin{figure}
\centering
\includegraphics[width=85mm]{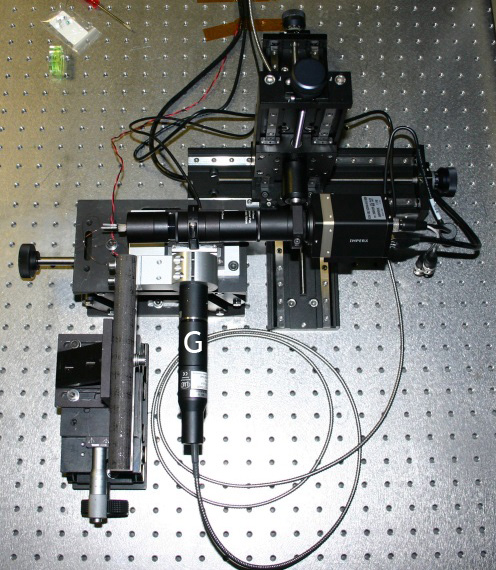}
\caption{Experimental set-up to scan a spot across the detector.
Bird's eye view. Key: G: confocal probe.}
\label{fig:lab_setup2}
\end{figure}

The Imperx camera and baffle tube were mounted on a manual $XYZ$ translation stage to allow the spot position to be moved coarsely around the CCD surface, and to adjust the camera position for best focus at different LED wavelengths.
The camera was rotated by $\sim$$30^\circ$ with respect to the piezo stage such that the image of the pinhole moved across the camera pixels in both $X$ and $Y$ simultaneously. 
Typically the pinhole source was moved by $5,000\times0.0333\,\mu$m or $20,000\times0.0111\,\mu$m steps and a light plus dark image pair collected at each position.
This step size was chosen to be a non-integer fraction of the pixel size. 

The GEV software delivered with the camera was used to capture images on a Windows PC. 
The software was configured for software triggering. 
Exposure times were adjusted to ensure a peak pixel count in the recorded images near $50\%$ of full-well capacity, i.e.\ $\sim$$2000$\,ADU (Analog to Digital Units) and avoid saturation. 
The Winger ultra-bright $5\,$mm LEDs used provided very high signal levels even at $5\,$mA drive current through a $5\,\mu$m pinhole. 
Consequently, exposure times were often as short as $3\,\mu$s.
Additional software written in Python: turned the LED on or off via a USB port; stepped the position of the piezo stage via a serial port; collected measurements from the confocal probe via a second serial port; captured images via {\sc mouseevent} function calls to ÔpressÕ the software trigger button in the GEV camera software; logged the stage and probe positions to a {\tt csv} file.
Images were saved as 12-bit non-normalized TIFF files, using $128\times128\,$pixel windows to minimise the amount of recorded data. 

Data were collected with a $472\,$nm LED and various pinhole diameters (1$\mu$m, 2$\mu$m, 5$\mu$m, 15$\mu$m, 50$\mu$m), plus $516\,$nm, $624\,$nm and $850\,$nm LEDs with the 5$\mu$m diameter pinhole.

\subsubsection{Model and control data}

For each test run, we predicted the shape of the spot using Zemax geometrical optics (figure~\ref{fig:lab_setup_diagram}) to model a point source through the matched achromat lens pair at optimized focus, and convolving with a circular top hat of the pinhole diameter.

We also simulated a control set of mock data in which the detector had perfectly uniform response. 
{To do this, we used software developed by \cite{imcom} to analytically place an Airy disc spot (with an appropriate range of sizes) at $5,000$ stepped locations with sub-pixel precision, before integrating it within pixels and realizing it with a flux of $\sim$$2000\,$ADU in the peak pixel.
We shall analyze this control data in the same way as our real data, to test the validity of our measurement methods in different regimes (it produces the solid lines in figures~\ref{fig:PRNUflux} and \ref{fig:PRNUposition}).}

\subsection{Data analysis}
\label{sect:cent-sw}

\subsubsection{Spot size measurement}
\label{sect:size-measure}

For large spots that are well resolved by the pixels, we measure their size directly from the images.
To do this in an automated way on noisy data, we measure the full width at half maximum (FWHM) of the best-fit Gaussian, optimizing the center, flux and size via Levenberg-Marquardt minimisation.
On the noiseless Zemax models, the FWHM of the best-fit Gaussian is typically within 4\% of the FWHM of the full spot profile.
After deconvolving from the pixel square, the measured FWHM is also within 6\% of the Zemax model, for all but the largest pinholes (where the true shape of the spot is probably influenced by structure in the LED).
For each test run, we record the mean FWHM measured in every exposure.

For small spots that are not well-resolved by the pixels, most of the light can fall within a single pixel.
The best-fit Gaussian then just fits the size of a pixel, and direct measurement becomes impossible.
For the test runs using a $472$\,nm LED and 1$\mu$m, 2$\mu$m, or 5$\mu$m pinholes, we use the FWHM of the best-fit Gaussian to the Zemax model.
However, we indicate up to 10\% uncertainty because of the difficulty focussing a spot that is not resolved by the detector and below the resolution limit of the lens.

\subsubsection{Spot position measurement}
\label{sect:cent-rrg}

To measure the observed position of spots, we adopt an iterative centroiding technique long-established for the high precision measurement of galaxy morphologies \citep{ksb95,rrg}. 
We initially guess the center of the brightest pixel to be the center $\bx_\mathrm{obs}^{i=0}$ of a spot image $I(\bx)$.
To improve this guess, we compute the image's first moments, weighted by a Gaussian centered at the current guess, and iterate
\be
\bx_\mathrm{obs}^{i+1}=\frac{\iint\mathrm{e}^{-|\bx-\bx_\mathrm{obs}^i|^2/(2\sigma_w^2)}~I(\bx)
~\bx~\rmd^2\bx}
{\iint\mathrm{e}^{-|\bx-\bx_\mathrm{obs}^i|^2/(2\sigma_w^2)}~I(\bx)
~\rmd^2\bx} \label{eqn:ksb}
\ee
where $\sigma_w$ is a weight size (see below), and the integral is performed over all pixels within a radius of $5\sigma$.
For our images, the iteration converges to a precision better than 0.001 pixels within 2 (3, 4) steps for stars inside the central 1\% (central 10\%, all) of the pixel.
In these tests, for which the calculations are not time-sensitive, we always perform 5 iterations and render this effect completely negligible.

To ensure a fair comparison, we fix the size of the weight function $\sigma_w$=2\,pixels.
The optimal choice of $\sigma_w$ for fastest convergence depends upon the size and S/N of the spot, but we obtain consistent results for all but pathologically small values of $\sigma_w$.
Note that $\sigma_w\!\rightarrow\!\infty$ recovers an unweighted centroid, but using this requires nonlinear thresholding to counteract the effect of noise \citep{ruf03,li09}. 

We also record the centers of both circular and elliptical Gaussians fitted to the spot images. 
This is similarly accurate as the moment-based method on simulated resolved spots with high S/N, but is slow and less robust to noise.
The mean offset between centroids calculated via moments and elliptical Gaussians in $20,000$ real exposures is $(13.1\pm0.1,-4.6\pm0.1)\times10^{-3}\,$pixels.
Of most concern is that neither algorithm works for spots that are smaller than a pixel.
Accurate fitting of very small spots requires precise knowledge of the spot wings \citep{tre03}.
In principle, it could be possible to shift and coadd our images to super-resolve the spot, then measure its shape \citep{imcom}.
However, this situation will not be encountered in the real instrument; and since our data contains only a single spot, applying sub-pixel shifts in order to measure the sub-pixel position is circular logic.
We therefore merely quantify the limitation of our algorithm, using simulated data where we know the true subpixel position.

\subsubsection{Spot flux measurement}
\label{sect:flux-measure}

Our experimental setup has greatly simplified measurements of the total flux within a spot, even for unresolved spots. 
We sum the flux in all pixels whose centers lie within an aperture of radius 5\,pixels or $2\times$FWHM (i.e.\ $4\times$ the half width at half maximum) of the spot centroid after convolution with the pixel square, whichever is larger. 
Our final results, which compare average measurements from thousands of images, are robust within 2\% for different choices of aperture radius.

\subsubsection{Input spot positions}
\label{sect:cent-refine}

The input position of a spot $(x_\mathrm{true},\,y_\mathrm{true})$ is set by the position of the translation stage.
We assume that the translation stage moves the optics on a straight line, the distance along which is measured up to $20,000$ times by the confocal probe as $\lconfocal(t)$.
We then obtain the 2D input position 
\ba
x_\mathrm{true}(t)=x_\mathrm{start}+g\lconfocal(t)\cos{\theta} ~~~~~~~~~~~~~ \nonumber \\
+\lspline(t)\cos{\theta}+x_\mathrm{spline}(t)\, \label{eqn:xtrue}  \\
y_\mathrm{true}(t)\,=y_\mathrm{start}\,+g\lconfocal(t)\sin{\theta} ~~~~~~~~~~~~~ \nonumber \\
+\lspline(t)\sin{\theta}+y_\mathrm{spline}(t)\, \label{eqn:ytrue}
\ea
by fitting the following free parameters, assuming initially that $(x_\mathrm{true},\,y_\mathrm{true})\!\approx\!(x_\mathrm{obs},\,y_\mathrm{obs})$:
\begin{itemize}
\item Starting subpixel position $(x_\mathrm{start},y_\mathrm{start})$ of the first exposure.
\item Angle $\theta$ between the spot scan line and the $x$-axis of the pixel grid.
\item Pitch of the steps $g$ between subsequent exposures, calibrating any linear gain in the confocal probe.
\end{itemize}
Plus, optionally:
\begin{itemize}
\item A cubic B-spline $\lspline$, calibrating low spatial frequency, non-linear gain in the confocal probe or judders in the translation stage.
\item Cubic B-splines $x_\mathrm{spline}$ and $y_\mathrm{spline}$ to perturb the true position of the spot at low spatial frequencies in the $x$ and $y$ pixel directions, accounting for residual vibration in the optical assembly, which needed to move, stop, and come to rest in between each exposure.
\end{itemize}
When we include the splines, we are careful to model variations only on scales larger than a pixel, to avoid spuriously introducing (or correcting) the kind of sub-pixel variations that we seek.

For a typical run with a $472\,$nm LED and $5\,\mu$m pinhole, the best-fit value for $g$ is $1+(4.0\pm0.2)\times10^{-3}$.
The best-fit $\lspline$, $x_\mathrm{spline}$, and $y_\mathrm{spline}$ have an rms of $74\,$nm, $4.6\,$nm, and $6.7\,$nm respectively.
The confocal probe had confirmed that our translation stage was juddering at around this level, so the fit is not surprising, but this demonstrates that our laboratory equipment is sufficiently stable to test the flight requirements.
For the simulated data of a perfect detector (and experiment, with $472\,$nm LED and $5\,\mu$m pinhole), the best-fit splines have an rms of $1.2\,$nm, $0.9\,$nm and $0.7\,$nm, and are consistent with zero everywhere.
This therefore demonstrates that the optional refinement is not introducing statistically significant perturbations.

\subsubsection{Modelling PRNU variation in 2D}

\begin{figure}[t]
\centering
\includegraphics[width=85mm]{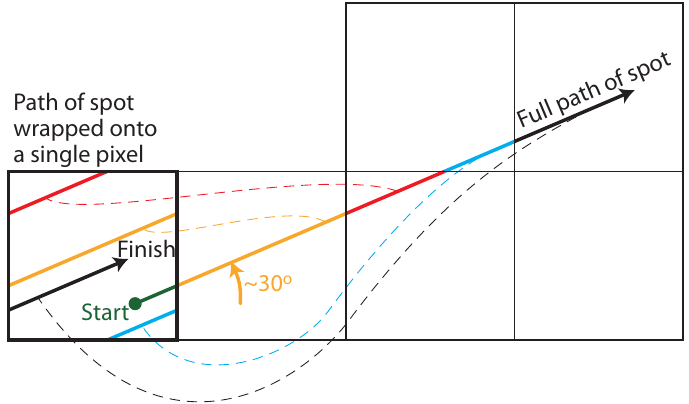}
\caption{We scan a spot in a continuous 1D line across several pixels.
As justified in the text, we then assume that sub-pixel PRNU is identical within small patches of the detector.
We then map the input spot positions back onto a single virtual pixel, so the entire 2D pattern of PRNU is well sampled.}
\label{fig:pixel_wrapping}
\end{figure}

We have not scanned a spot in a 2D pattern around one pixel, but in a 1D pattern across several pixels.
To map 2D PRNU, we assume that any pattern is repeated identically in all the pixels within small ($\sim\!25\!\times\!15$\,pixel) patches of the detector.
We then wrap the sub-pixel input spot positions such that $0\!<\!x_\mathrm{true}\!<\!1$ and $0\!<\!y_\mathrm{true}\!<\!1$, mapping them all back onto a virtual pixel that becomes well sampled (figure~\ref{fig:pixel_wrapping}).

To check the validity of this assumption, we have:
(i) performed the test runs (including some repeat runs with otherwise identical configuration) in different parts of the detector; the results from these runs are consistent with each other, within experimental errors.
(ii) explicitly tried to fit patterns across $2\times2$ groups of pixels. It might be possible to have such patterns if e.g.\ a chip mask used to manufacture devices with an RGB Bayer mosaic were adapted to create this monochromatic device. 
We find that power in a 2-pixel pattern is less than 1\% of that in the 1-pixel pattern, and consistent with zero.
We therefore assume that the tiny residuals on these spatial scales are due to vibrations in the translation stage, and allow them to be fitted away by the B-splines in equations~\eqref{eqn:xtrue} and \eqref{eqn:ytrue}.

\subsection{Results}
\label{sect:cent-results}

\subsubsection{Spot sizes}

The sizes of the spot produced by different combinations of LED wavelength and pinhole size are listed in table~\ref{tab:spotsize}. Colors indicative of the LED wavelengths are used for the data points in figures~\ref{fig:PRNUflux} and \ref{fig:PRNUposition}.

\begin{table}[t]
\caption{Sizes of the spot produced with various experimental configurations and scanned across the detector. Measurements from a Zemax model or real images are consistent within $6$\% (see Section \ref{sect:size-measure}).  \label{tab:spotsize}}
{\scriptsize
\begin{center}
    \begin{tabular}{lrr@{.}l}
    \hline
    LED wavelength [nm] & Pinhole diameter [$\mu$m] & \multicolumn{2}{c}{Spot FWHM [$\mu$m]}  \\ \hline
~~~~~472 (blue) & 1~~~~~~~~~~~~~~~~~ & ~~~~~~~~~3 & 57 \\
~~~~~472 & 2~~~~~~~~~~~~~~~~~ & 3 & 86 \\
~~~~~472 & 5~~~~~~~~~~~~~~~~~ & 5 & 05 \\
~~~~~516 (green) & 5~~~~~~~~~~~~~~~~~ & 7 & 02 \\
~~~~~624 (red) & 5~~~~~~~~~~~~~~~~~ & 12 & 44 \\
~~~~~472 & 15~~~~~~~~~~~~~~~~~ & 14 & 35 \\
~~~~~850 (near-IR) & 5~~~~~~~~~~~~~~~~~ & 17 & 06 \\
~~~~~472 & 50~~~~~~~~~~~~~~~~~ & 42 & 74 \\
    \hline
    \end{tabular}
\end{center}
}
\end{table}

\subsubsection{Flux loss as a function of sub-pixel position}

\begin{figure}[t]
\centering
\includegraphics[width=80mm]{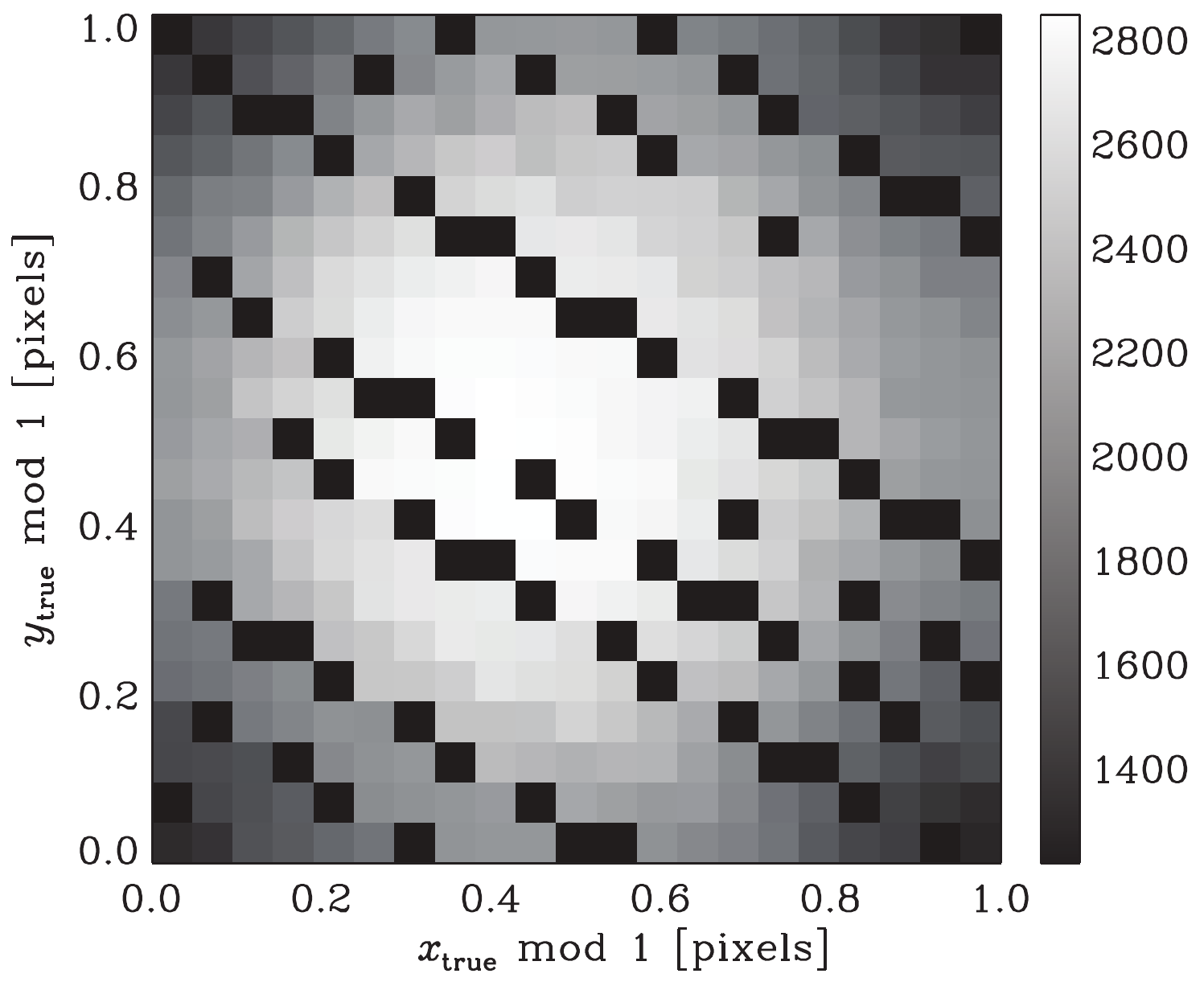}
\caption{Pixel Response Non-Uniformity: the total flux recorded (within a 5\,pixel aperture) when our smallest spots ($1\,\mu$m pinhole, $472$\,nm LED) are shone at different sub-pixel positions. Black regions contain no data, and the scanning strategy of figure~\ref{fig:pixel_wrapping} is apparent. The response of a perfect detector would be uniform, but here some flux is lost near the edges of pixels.}
\label{fig:PRNUfluxgrid}
\end{figure}

For very small spots, the variation in observed total flux at different sub-pixel positions $\Delta F(x_\mathrm{true},y_\mathrm{true})$ is shown in figure~\ref{fig:PRNUfluxgrid} and well fit by 
\ba
\Delta F = F_\mathrm{peak} \big[ 1\!-\!A_{x}\sin^4\!{\left(\pi (x_\mathrm{true}-x_c){/1\,\mathrm{pixel}}\right)} ~\,\nonumber \\
-\!A_{y}\sin^4\!{\left(\pi (y_\mathrm{true}-y_c\,){/1\,\mathrm{pixel}}\right)} \big]\!. \label{eqn:deltaf_def}
\ea
This pattern peaks at the center of the pixel, $(x_c,\,y_c)$, and drops at the sides in a continuous way between adjacent pixels; it thereby mirrors the likely shape of the microlens array focussing light onto the responsive areas of the CCD.
The exponent 4 represents a compromise between values that best fit data from test runs with various spot sizes; this varies between 3--5 if left as a free parameter, but we fix it to simplify comparison between runs.

\begin{figure}[t]
\centering
\includegraphics[width=85mm]{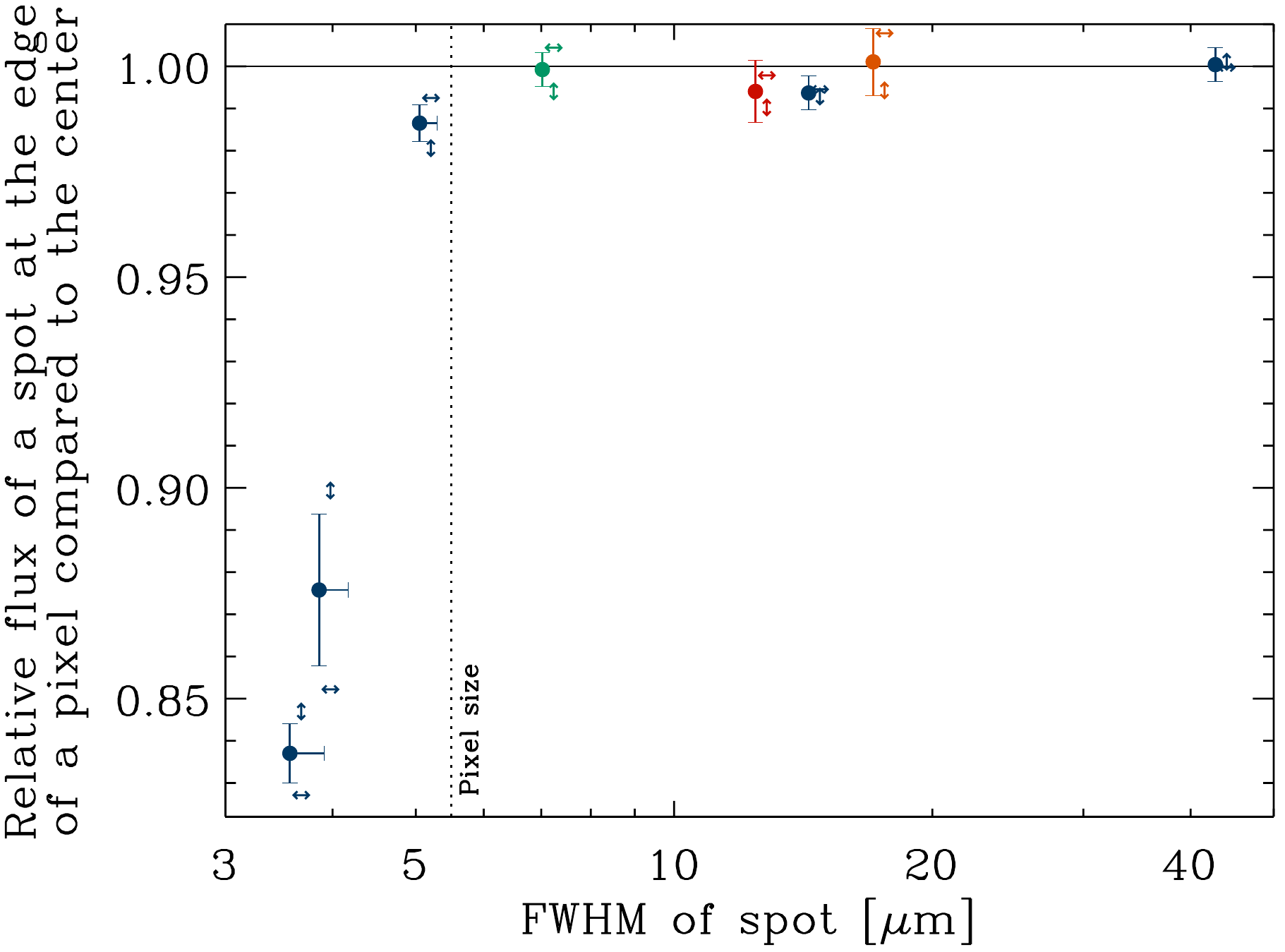}
\caption{Pixel Response Non-Uniformity: the fraction of flux lost at the edges of a pixel, from an input spot of varying widths.
Data points show $1-(A_{x}+A_{y})/2$, which are defined in equation~\eqref{eqn:deltaf_def}.
Offset slightly to the right {and without error bars} for clarity, the left-right arrows show $1-A_{x}$ and the up-down arrows show $1-A_{y}$ individually.
The leftmost data corresponds to figure~\ref{fig:PRNUfluxgrid}.
A perfect detector would have $A_{x}=A_{y}=0$.}
\label{fig:PRNUflux}
\end{figure}

For test runs with every size spot, we determine best-fit values for the free parameters $F_\mathrm{peak}$, $(x_c,\,y_c)$, and $A_{x}$, $A_{y}$ using Levenberg-Marquardt minimisation.
Measured values for $A_{x}$ and $A_{y}$ are shown in figure~\ref{fig:PRNUflux}; a perfect detector would have $A_x=A_y=0$, as indeed recovered in our mock test data with simulated images.

The flux loss detected for small spots is therefore real and significant.
The greater flux loss at the sides of a pixel than at the top/bottom ($A_x\!>\!A_y$) is {also expected, due to the asymmetric detector architecture in which the responsive photodiode and} non-responsive interline readout columns lie to the sides of the pixels.
{Indeed, the best-fit value for $(x_c,\,y_c)$ in figure~\ref{fig:PRNUfluxgrid} is offset laterally to $(0.468\pm0.003,\,0.502\pm0.003)\,$pixels, although this result varies slightly with position within the detector, whose microlenses are positioned in a telecentric pattern.}
Our results thus indicate that some light is still lost into the nonresponsive regions, despite the microlens array. 

However, the observed flux in large spots that are resolved by the pixels (especially if they are Nyquist sampled or better) are unaffected by this PRNU within our experimental precision.
Some flux loss may indeed contribute to the overall QE (here degenerate with $F_\mathrm{peak}$) but, with the loss averaged over many {locations}, the flux response appears uniform for a spot positioned anywhere.

\subsubsection{Centroid shift as a function of sub-pixel position}

\begin{figure}[t]
\centering
\includegraphics[width=85mm]{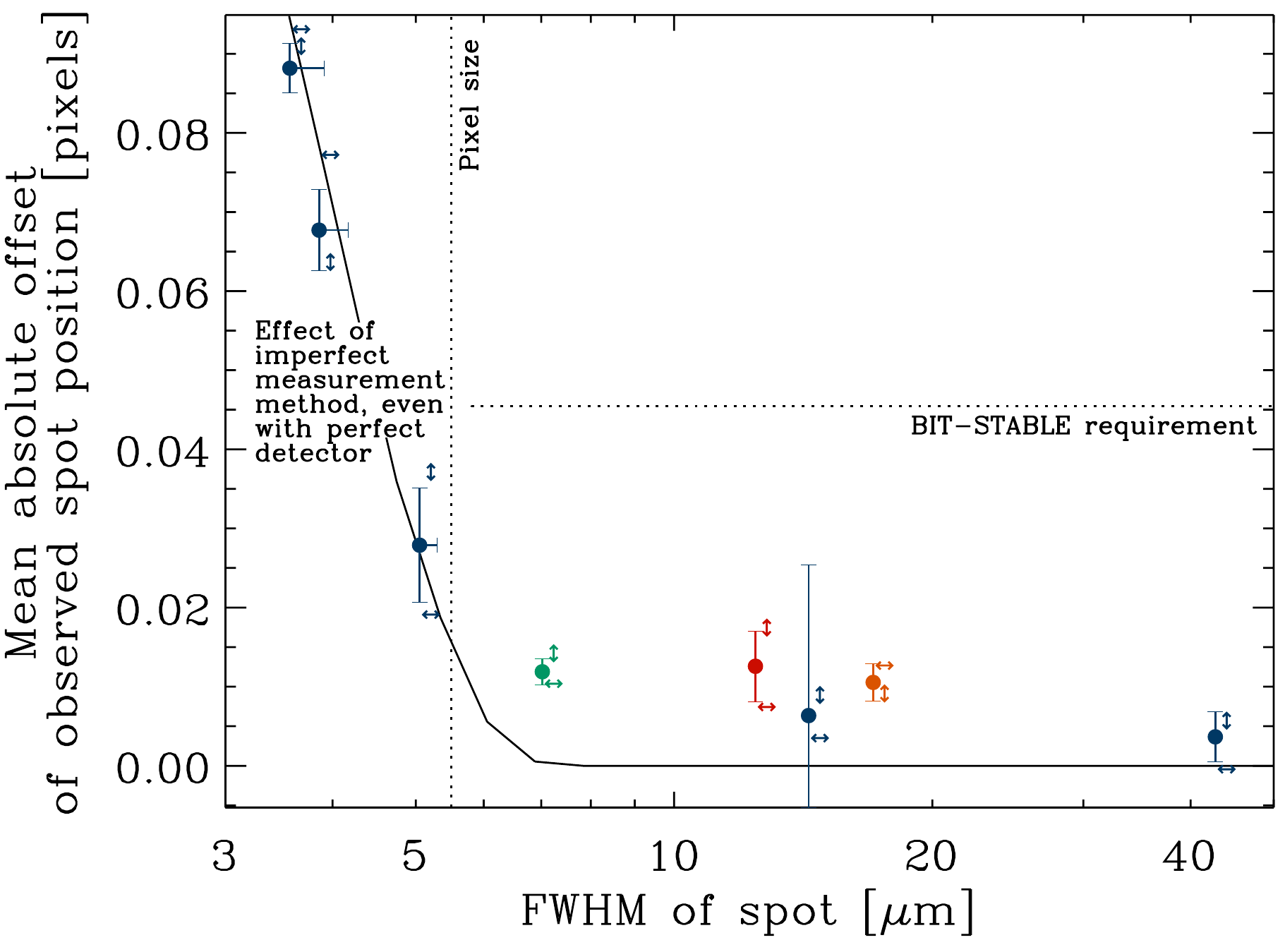}
\caption{Pixel Response Non-Uniformity: the spurious shift in the position of an input spot of varying widths at the edges of a pixel.
Following equation~\eqref{eqn:deltaf_def}, data points show $(B_{x}+B_{y})/2$.
Offset slightly to the right {and without error bars} for clarity, the left-right arrows show $B_{x}$ and the up-down arrows show $B_{y}$ individually.
A perfect detector would have $B_{x}=B_{y}=0$, and BIT-STABLE requires performance below the horizontal line.}
\label{fig:PRNUposition}
\end{figure}

The centroid of a spot can be shifted both by a gradient in detector sensitivity across the spot, or (more likely) by optical distortion through the lenslet array.
{Once again reflecting the lens geometry, we fit offsets in the observed centroid compared to its input position $(\Delta x,\,\Delta y)\equiv(x_\mathrm{obs}\!-x_\mathrm{true},\,y_\mathrm{obs}\!-y_\mathrm{true})$ as it moves within a pixel as}
\ba
\Delta x = B_{x}\sin{(2\pi (x_c-x_\mathrm{true}){/1\,\mathrm{pixel}})} \, \label{eqn:deltax_def_simple} \\
\Delta y = B_{y}\sin{(2\pi (y_c-y_\mathrm{true}){/1\,\mathrm{pixel}})} . \label{eqn:deltay_def_simple}
\ea
This pattern is zero in the middle of the pixel and at the edges, where symmetry also ensures that the lenslet must also be flat.
Nuisance parameters $(x_c,\,y_c)$ are degenerate with $(x_\mathrm{start},\,y_\mathrm{start})$ in equations~\eqref{eqn:xtrue} and \eqref{eqn:ytrue}, but are left as free parameters here for completeness.
Positive values of $B_x$ or $B_y$ indicate a shift towards the center of a pixel.
This function, including the (lack of) exponent, is a very good fit to the behavior of an imperfect position measurement algorithm when applied to simulated images from a perfect detector (see \S\ref{sect:cent-rrg}).
To try to dig out any other signal, we have also tried varying the exponent, or doubling the frequency of the sinusoidal variations.
However, no other significant signal is found.
For example, for the test run using the $472\,$nm LED and $5\,\mu$m pinhole, the best-fit exponent is $1.01\pm0.17$, the amplitudes $B_x$, $B_y$ are unaffected within $1\%$ by the addition of extra terms with double frequency, whose best-fit amplitudes are less than $3\%$ of $B_x$ and $B_y$.

The mean absolute offsets, $\sqrt{\Delta x^2+\Delta y^2}$, averaged within an entire pixel, are shown for different sized spots in figure~\ref{fig:PRNUposition}.
The difficulty of position measurements for small spots makes it hard to distinguish between their expected centroid shifts and imperfect algorithmic behavior with any statistically precision.
For a Nyquist-sampled spot, the mean offset is around $0.01$\,pixels=$55$\,nm.
For a large spot, the mean offset appears to tend to zero, as expected if the individual shifts within many pixels average away.
The mean offset is the appropriate quantity to compare to requirements because noise in the position measurement from an individual image with a guide camera will be dominated by photon-counting shot noise.
The {\em peak} offset anywhere within the pixel is $\sim\!1.5$ times greater than the mean offset but even this, for a Nyquist-sampled PSF, is still within requirements for BIT-STABLE.

\section{Windowed readout} \label{sec:window}

\subsection{The problem}

The BIT-STABLE mission requires 
{$>\!100\!\times\!100$\,pixel images of a guide star at $>\!50$\,Hz.
Because of the finite readout speed and data transfer rate over 100Mbps ethernet, it will not be possible during flight to read out the full frame and crop the region of interest in postprocessing.}
During further camera testing at the Jet Propulsion Laboratories, California Institute of Technology (JPL), we noticed {artifacts in} images obtained using Windowed Readout Mode.
Compared to images obtained using Full Frame Readout Mode, we observed an {apparent} overall loss of sensitivity, and a spatially varying sensitivity gradient in the readout direction.

Any loss of CCD sensitivity will degrade photometric measurements, and raise the magnitude limit of guide stars for the BIT-STABLE mission. 
{Uncorrected} spatial variations in sensitivity will also affect the camera's measurements of (relative) position.
The effect can be computed by inserting a {multiplicative} factor $S(\bx)$ into the numerator and denominator of equation~\eqref{eqn:ksb}.
Let us assume that the PSF image $I(\bx)$ is a Gaussian of width $\sigma_p$, centered at $\bx_\mathrm{true}\!=\!\bx_\mathrm{o}+{\mathbf\delta}\bx$, where $\bx_\mathrm{o}$ is a nearby reference point. 
For a sensitivity variation in the readout direction expressed as a Taylor series about $\bx_\mathrm{o}$
\be
S(\bx)=a+b(y-y_\mathrm{o})+c(y-y_\mathrm{o})^2, \label{eqn:staylor}
\ee
the observed $y$ position of the PSF image becomes
\ba
y_\mathrm{obs} & = &
y_\mathrm{true}+
\frac{\sigma^2(b+2c\delta y)}{a+b\delta y+c\delta y^2+c\sigma^2} \\
& \approx &
y_\mathrm{true}+
\frac{b}{a}\sigma^2+\frac{2c}{a}\sigma^2\delta y \label{eqn:offsets}
\ea
to first order in $\delta y$ and leading order in $\nicefrac{b}{a}$ and $\nicefrac{c}{a}$, and where $\sigma^2\!=\!\sigma_p^2+\sigma_w^2$. 

A constant sensitivity gradient would thus shift all measured positions by a constant offset -- which would be unnoticeable in 
either guide star imaging (where only the relative position between exposures is measured) or our spot projection tests (where we fitted the absolute position $\bx_\mathrm{start}$ as a free parameter).
Only second order (and higher) spatial derivatives of sensitivity affect measurements of the offset between two sources, or a moving source observed in successive exposures, via
\be
\frac{\rmd y_\mathrm{obs} }{\rmd y_\mathrm{true}}=1+\frac{2c}{a}\sigma^2.
\ee

\subsection{Experimental setup}

This testing was performed at the Jet Propulsion Laboratories (JPL).
The camera was exposed to a near-flat DC light source, and images were captured in both Full Frame and Windowed Readout Mode (under the same lighting conditions, camera modes, and exposure times).

\subsection{Results}

{\subsubsection{Observed behavior}}

\begin{figure}[t]
  \centering
  \includegraphics[width = 0.5\textwidth]{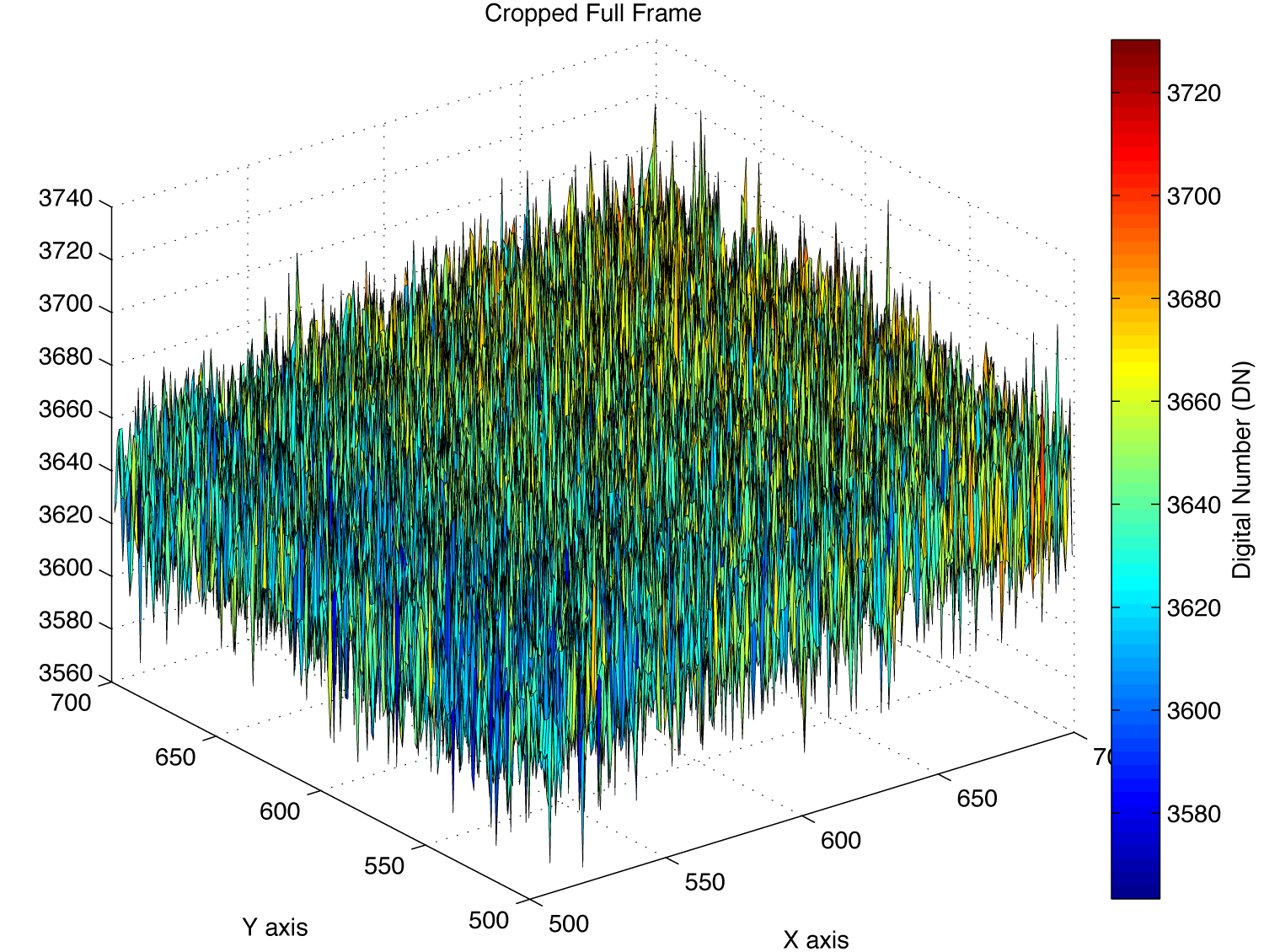}
  \caption{Flat field image obtained in Full Frame Readout Mode.
  A cropped, $200\times 200$\,pixel region is shown.
  To reduce readout and shot noise, the image has been averaged over 35 consecutive exposures.}
  \label{fig:cropped_full_35}
\end{figure}

\begin{figure}[t]
  \centering
  \includegraphics[width = 0.5\textwidth]{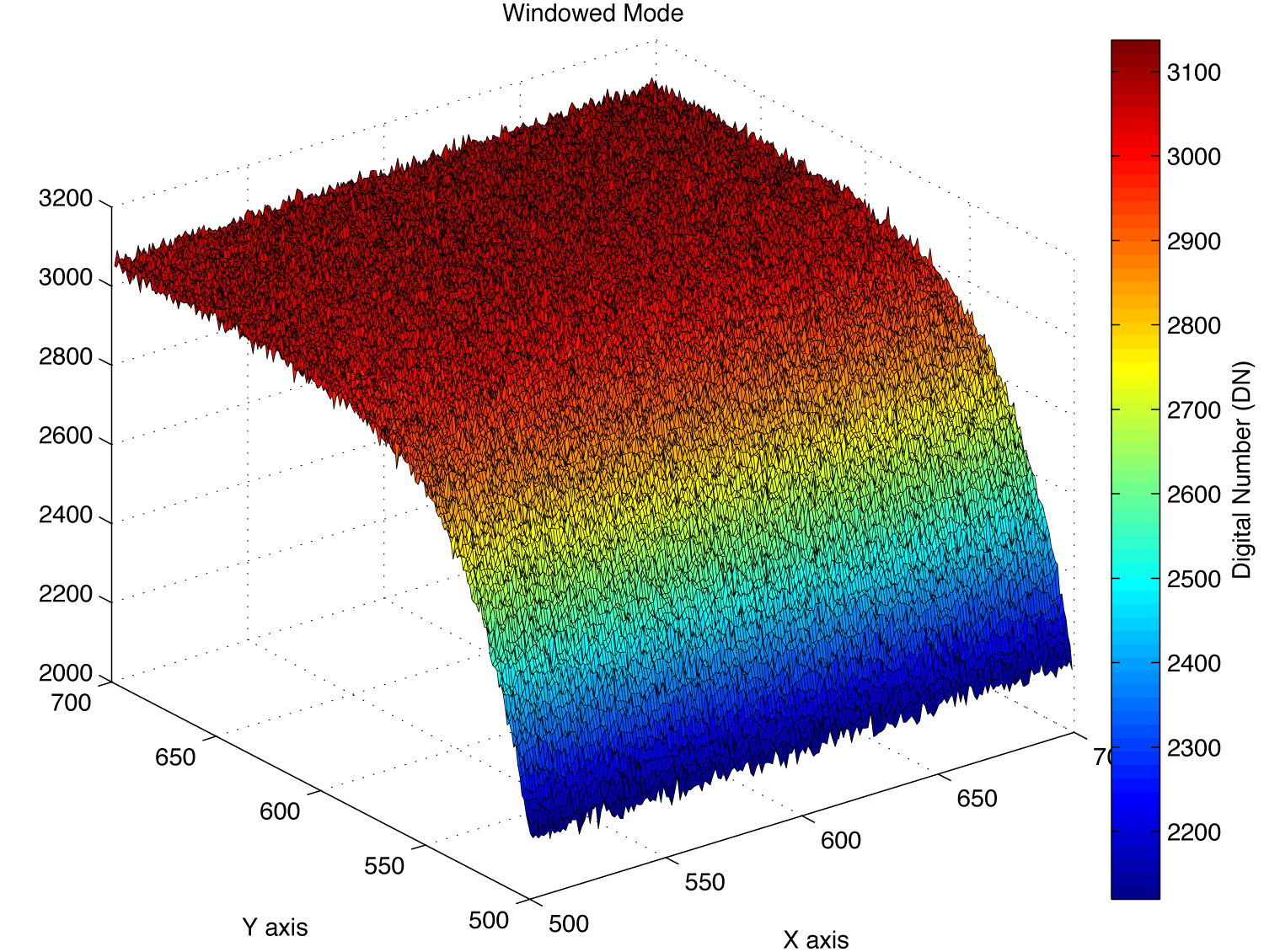}
  \caption{Flat field image obtained in Windowed Mode.
  The $200\times 200$\,pixel window corresponds to the same pixels as those shown in figure~\ref{fig:cropped_full_35}.
  To reduce readout and shot noise, the image has been averaged over 35 consecutive exposures.}
  \label{fig:windowed_35}
\end{figure}

In Full Frame Readout Mode, the received {flat field} image is roughly flat, as expected.
Figure~\ref{fig:cropped_full_35} shows pixels $(500$:$700,500$:$700)$ cropped from full frame images, averaging over 35 consecutive exposures to reduce the amount of readout and shot noise\footnote{{By considering flat field images with different exposure times, we find that the readout noise and dark current are consistent with the $12\,$e$^\mathrm{-}$~rms and $1\,$e$^\mathrm{-}\!/$s values quoted in the detector specifications \citep{truesense}.}}.
However, the same $200\!\times\!200$\,pixels captured in Windowed Mode exhibit a reduced signal and a ``waterfall''-like gradient (figure~\ref{fig:windowed_35}).
To characterize the drop in sensitivity, we divide the Windowed Mode image (figure~\ref{fig:windowed_35}) by the corresponding pixels of the Full Frame Readout Mode image (figure~\ref{fig:cropped_full_35}).
Spatial variation in the result is well fit by
\ba
S(\bx)=(0.841\pm0.001)~~~~~~~~~~~~~~~~~~~~~~~~~~~\nonumber\\
-(0.253\pm0.001)\exp\!{\left(\frac{500-y}{28.8\pm0.18}\right)}. \label{eqn:sfit}
\ea
The $84.1\%$ drop in sensitivity is constant within a few percent if the window is moved around the CCD.
The drop is negligible for the largest possible (e.g.\,$1700\!\times\!2000$\,pixel) windows, and becomes progressively worse as the window size is reduced. 

\begin{figure}
  \centering
  \includegraphics[width = 0.5\textwidth]{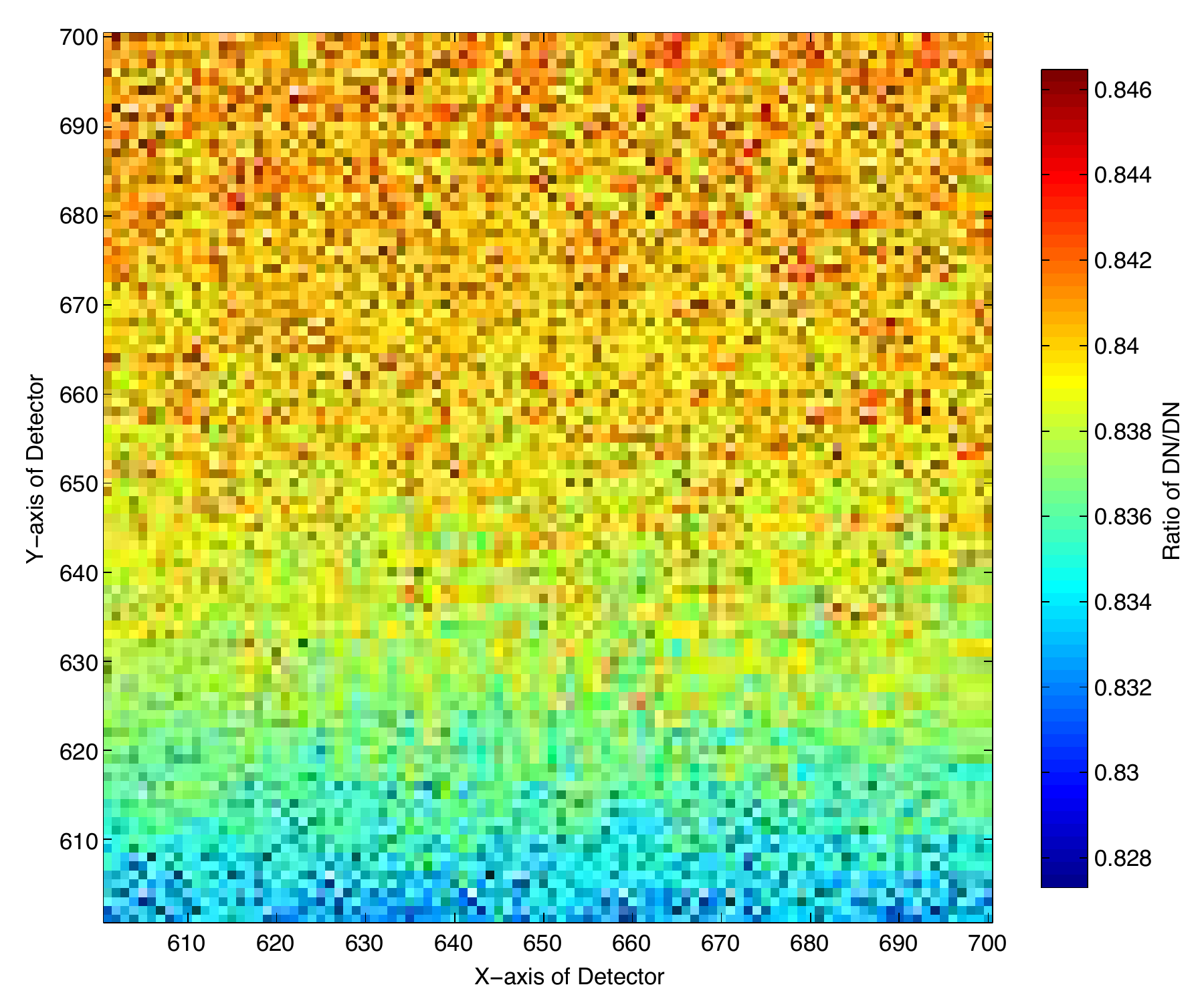}
  \caption{The net effect on sensitivity, when using our workaround of obtaining a larger window than required.
  This shows the ratio of counts in the Windowed Mode image compared to those in the Full Frame Readout Mode image, in the `flat' $100\!\times\!100$\,pixel region at the top of the window.}
  \label{fig:scaling_window}
\end{figure}

{Dark frames observed in Windowed Readout Mode also exhibit a spurious excess signal.
An additive bias of $\sim\!1\,$ADU is present in the first row, decreasing linearly in the readout direction to $\sim\!0.1\,$ADU after 50 rows.
The spurious signal alternates by $\pm10\%$ on odd/even columns.
However, it is constant, regardless of the exposure time.}

{\subsubsection{Possible explanation}}

The camera manufacturer reported that the sensor does not have a hardware fast {line} dump{, which is used in other devices to quickly discard charge from all pixels in a row and rapidly advance readout toward a restricted region of interest.
To produce windowed readout from the IGV-B2320-M, this feature is being emulated.}
The exponential form of equation~\eqref{eqn:sfit} is {reminiscent of charge transfer problems \citep[e.g.][]{cti1,cti2}, perhaps because the clock speed is being pushed higher than the $40\,$MHz maximum specification \citep{truesense}.
The loss of sensitivity may then be caused by a buildup of (undumped) charge, which affects the sensitivity of the output amplifier.}
It would be useful to compare this camera's performance with that of the IGV-B2020-M, which utilises the Truesense KAI-04022 detector ($2048\!\times\!2048$ $7.4$\,$\mu$m pixels) and which does have a fast {line} dump.

Note that if the cause of sensitivity loss is indeed undumped charge, a flat field with a high level of charge outside the window presents a worst case scenario.
Images from a guide camera that contain mostly blank sky should be less affected.
It would be interesting to make photometric measurements of an isolated spot {of constant flux as it moves a long distance across the detector} in Windowed Readout and Full Frame Readout Modes.
However, our spot projector lab is not 
currently
available for further tests of this camera.

{\subsubsection{Mitigation strategy}}

{The low level of spurious additive signal can be removed via subtraction of dark frames obtained in flight.
Multiplicative sensitivity variation could be reduced via a flat field.
However, in case the effect is not constant, or signal-dependent, we require a more robust mitigation strategy.}

If the first $100\!\times\!100$\,pixel region to be read out were adopted{, without any correction,} as the region of interest (as it was in our spot projection tests of section~\ref{sec:cent}), it would have a mean sensitivity of $0.770\pm0.003$ and a steep local gradient in sensitivity.
Approximating equation~\eqref{eqn:sfit} as a Taylor series near the middle of this region (i.e.\ with $y_\mathrm{o}\!=\!550$) yields 
$a=0.796$, $b=1.55\!\times\!10^{-3}$\,pixel$^{-1}$ and $c=-5.37\!\times\!10^{-5}$\,pixel$^{-2}$.
With a Nyquist sampled PSF, for which $\sigma_p=2/2.3548$\,pixels and hence $\sigma\!=\!2.2$\,pixels, equation~\eqref{eqn:offsets} suggests that $|\rmd y_\mathrm{obs}/\rmd y_\mathrm{true}-1|\approx6.4\times10^{-4}=3.5$\,nm/pixel. 
{This would formally fail mission requirements if the spot frequently moved by $100\,$pixels between successive exposures (but the closed-loop guidance would converge so long as typical movements are smaller).
Across the $\sim\!15$ pixel rows traversed in our spot projection tests, this large-scale variation could explain a fraction of the $74\,$nm rms fluctutation of} $\lspline$ that we had previously ascribed to juddering of the translation stage.

As a workaround to further militate against sensitivity loss and its spatial variation, we capture a taller window than is necessary, then discard the first 100 rows and retain only the region farthest from the readout register -- in which the sensitivity is higher and approximately flat (figure~\ref{fig:scaling_window}).
For example, if the region of interest is pixels $(600$:$700,600$:$700)$, we capture pixels $(600$:$700,500$:$700)$.
At 50\,Hz frame rate, this $100\!\times\!200$\,pixel window remains well within the maximum ethernet transfer rate (see section~\ref{sec:gige}) and the camera's maximum frame rate (which depends only on the number of rows read out).
In the retained region, the mean sensitivity is $0.839\pm0.002$ and its spatial variation is well fit by equation~\eqref{eqn:staylor} with $y_\mathrm{o}=650$, $a=0.839\pm0.002$, $b=(6.88\pm5.36)\!\times\!10^{-5}$ pixel$^{-1}$, and $c=(-1.04\pm2.08)\!\times\!10^{-6}$\,pixel$^{-2}$.
Again assuming a Nyquist sampled PSF, this indicates that $|\rmd y_\mathrm{obs}/\rmd y_\mathrm{true}\!-\!1|\approx1.2\times10^{-5}=0.064$\,nm/pixel. 
This level of error will have a negligible effect on the device's performance as a fine guidance camera, so long as the input motion of the guide star between exposures keeps it within the 100\,pixel region of interest --- and the mission would fail anyway if it moved farther than that.\\

\section{Conclusions}
\label{sec:conc}

We have successfully demonstrated that an unmodified commercial off-the-shelf front-illuminated progressive scan interline transfer CCD guide camera can meet the requirements of the closed-loop star-guiding and tip-tilt correction system for the BIT-STABLE balloon mission.

The camera can withstand the $+70^\circ$C / $-40^\circ$C survivability and $+60^\circ$C / $-30^\circ$C operating temperature requirements in near-vacuum (pressures of 4mBar or below). 
Incident light has to pass through a microlens array and be focussed onto photoresponsive regions of silicon in order to be detected. 
Some light remains undetected, particularly at the sides of pixels adjacent to the interline readout columns.
However, this effect is averaged out over several pixels for a Nyquist-sampled spot, whose centroid can be measured to 1\% (1.5\%) of a pixel in an average (worst) case sub-pixel position.
artifacts in sensitivity and flat field are created in the first $\sim100$ rows of a windowed image because the detector does not have a hardware fast dump.
The required performance can be recovered by capturing larger images, and discarding the first 100 rows.

Data transfer from the camera works reliably in $100\,$Mbps Full Duplex ethernet mode, but not in GigE ethernet mode.
This lowers the theoretical maximum frame rate for full-frame images, but remains within the requirements of BIT-STABLE.

\section*{Acknowledgments}

The authors thank Barnaby Rowe for software to create pixellated PSF models, Carl Christian Liebe for guidance on noise analysis{, and Roger Smith for discussions of charge transport}.
We acknowledge support from Durham University's Centre for Advanced Instrumentation, Edinburgh University sustainability fund, UK Science and Technology Facilities Council (grant number ST/L00075X/1), European Research Council (grant number MIRG-CT-208994), and the Leverhulme Trust (grant number PLP-2011-003).
{This work was supported in part by the National Science Foundation under Grant No. PHYS-1066293 and the hospitality of the Aspen Center for Physics.}
RM is supported by a a Royal Society Research Fellowship.
The work of HC, LLJ, MM and SS was carried out at the Jet Propulsion Laboratory, California Institute of Technology, under contract with the National Aeronautics and Space Administration (NASA). 



%\appendix
%\section{} \label{sec:nospline}
%\clearpage

\end{document}